\documentclass[sigconf,nonacm]{acmart} %remove nonacm=true for final version. It is just to block the running titles from 2nd page onwards

\usepackage{array}
\usepackage{listings}

\usepackage{tabularx}
\usepackage{rotating}

\usepackage{color}

\newcommand{\boldtitle}[1]{\vspace{5px}\noindent\textbf{#1}}

\usepackage[justification=centering]{caption}

\usepackage{pifont}
\newcommand{\vuln}{\textcolor{red}{\ding{55}}}
\newcommand{\notvuln}{\textcolor{green}{\ding{51}}}
\newcommand{\na}{--}

\graphicspath{{./figures/}}

\begin{document}

\author{Thanh Bui}
\email{thanh.bui@aalto.fi}
\affiliation{%
  \institution{Department of Computer Science \\ Aalto Univeristy, Finland}
}

\author{Siddharth Prakash Rao}
\email{siddharth.rao@aalto.fi}
\affiliation{%
  \institution{Department of Computer Science \\ Aalto Univeristy, Finland}
}

\author{Markku Antikainen}
\email{markku.antikainen@aalto.fi}
\affiliation{%
  \institution{Department of Computer Science \\ Aalto Univeristy, Finland}
}

\author{Tuomas Aura}
\email{tuomas.aura@aalto.fi}
\affiliation{%
  \institution{Department of Computer Science \\ Aalto Univeristy, Finland}
}

\renewcommand{\shortauthors}{\center{\textcolor{blue} {*A \href{https://link.springer.com/chapter/10.1007/978-3-030-35055-0_7}{\textbf{refined version}} of this draft, with the same title, has been published in the 24th Nordic Conference on Secure IT Systems.}}}

\title{Client-side Vulnerabilities in Commercial VPNs}
\renewcommand{\shorttitle}{Client-side Vulnerabilities in Commercial VPNs}

%!TEX root = ../main.tex
\begin{abstract}
Internet users increasingly rely on commercial virtual private network (VPN) services to protect their security and privacy. The VPN services route the client's traffic over an encrypted tunnel to a VPN gateway in the cloud. Thus, they hide the client's real IP address from online services, and they also shield the user's connections from perceived threats in the access networks. In this paper, we study the security of such commercial VPN services. The focus is on how the client applications set up VPN tunnels, and how the service providers instruct users to configure generic client software. We analyze common VPN protocols and implementations on Windows, macOS and Ubuntu. We find that the VPN clients have various configuration flaws, which an attacker can exploit to strip off traffic encryption or to bypass authentication of the VPN gateway. In some cases, the attacker can also steal the VPN user's username and password. We suggest ways to mitigate each of the discovered vulnerabilities.
\end{abstract}




\maketitle

%!TEX root = ../main.tex

\section{Introduction}
\label{sec:intro}

Virtual private networks (VPN)~\cite{10290} were originally developed for connecting geographically distributed corporate networks to each other with encrypted tunnels, so that they would form a single secure logical network. Their functionality was also extended for connecting remote workers to the employer's intranet. However, one of the most common uses of VPN today is to protect regular Internet users who seek improved security and privacy. Such users perceive a need for a VPN in many different situations, such as when accessing the Internet over public Wi-Fi (e.g.~at a cafe, hotel, or airport), to hide their online activities from an oppressive entity (e.g.~government, employer or Internet service provider), or to access geoblocked media content. Because of the increased demand, a large number of \emph{commercial VPN services} have appeared in the market~\cite{vpnfuture}.

Commercial VPNs typically function by tunneling the user's Internet traffic through a trusted remote server before it is forwarded to its final destination. This achieves two goals: first, the traffic is protected by an encrypted VPN tunnel against dangers in the access network and, second, the destination server does not learn the real IP address of the client. There are quite a few VPN protocols that can be used to establish the tunnel (e.g.~PPTP, SSTP, OpenVPN), and the commercial VPN providers usually support several of them. The commercial VPN providers also provide native client applications with graphical user interfaces, which allow the user to select the protocol and server and set up the VPN connection accordingly. For more technically savvy users who prefer not to install the provided application, the VPN service providers usually give instructions for configuring the built-in VPN client in the user's operating system (OS) to work with their servers.

\begin{table*}[ht!]
    \centering
    \small
    \def\arraystretch{1.1}
    \begin{tabular}{c|c|c|c|c|c|c|c}
     % \hline

        & \textbf{PPTP}
        & \textbf{SSTP}
        & \textbf{L2TP/IPSec}
        & \textbf{Cisco IPSec}
        & \textbf{IKEv2}
        & \textbf{OpenVPN}
        & \textbf{SoftEther VPN}
        \\ \hline
    \textbf{Windows}
        & Built-in % PPTP
        & Built-in % SSTP
        & Built-in% L2TP
        & Shrew Soft % Cisco
        & Built-in % IKEv2
        & OpenVPN % OpenVPN
        & SoftEther % SoftEther
        \\ \hline
    \textbf{macOS}
        & \na % PPTP
        & EasySSTP % SSTP
        & Built-in % L2TP
        & Built-in % Cisco
        & Built-in % IKEv2
        & OpenVPN % OpenVPN
        & SoftEther % SoftEther
        \\ \hline
    \textbf{Ubuntu}
        & Built-in % PPTP
        & sstp-client % SSTP
        & xl2tpd % L2TP
        & Shrew Soft % Cisco
        & StrongSwan % IKEv2
        & OpenVPN % OpenVPN
        & SoftEther % SoftEther
        \\
        % \hline
    \end{tabular}
    \captionsetup{justification=centering}
    \caption{VPN clients that are used or recommended by commercial VPN providers.}
    \label{tab:vpn_clients}
\end{table*}

The commercial VPN services have undergone severe scrutiny~\cite{perta2015glance,appelbaum2012vpwns,CVE-2018-3952,CVE-2018-4010,al2017one,rossberg2011survey,portforwardingleak}, which has exposed various malpractices and vulnerabilities in the services. A handful of researchers have taken a closer look at the client configuration in the commercial VPNs~\cite{perta2015glance,appelbaum2012vpwns,al2017one}. These studies revealed a number of misconfigurations of popular VPN services that lead to user de-anonymization and traffic leakage. Our work extends this theme in the literature with the focus on the security of the VPN tunnels, namely whether they are encrypted and authenticated properly.

\boldtitle{Motivation. }
Our work was primarily motivated by the observation that many commercial VPN providers configure L2TP/IPsec, a popular VPN protocol, in an insecure way. Specifically, the protocol relies on IPsec~\cite{atkinson1998security} to provide the secure transport, but many VPN providers use one pre-shared key for all users to authenticate the IPsec tunnels. These service-specific keys are either publicly available online or can be discovered by examining the client configuration. An example of such a pre-shared key is ``12345678''. When an attacker knows the pre-shared key, it can perform a man-in-the-middle (MitM) attack on the VPN connection and, as the result, obtain all the network traffic to and from the victim's computer. This problem was already discussed on public forums in 2016~\cite{l2tpkeys}. When we re-analyzed the 14 insecure commercial VPN services mentioned in the discussion, we found that only four of them had fixed the problem or stopped supporting the L2TP/IPsec protocol, while 10 were still using the insecure configuration. Whatever the reason is, this security issue remains opaque to most end-users. Thus, we feel that it is important to scrutinize systematically the client configurations of commercial VPN services, considering both L2TP/IPsec and other protocols, for flaws that could undermine the user's security and privacy.

\boldtitle{Contributions. }
In this paper, we study how popular commercial VPN providers set up, or how they instruct users to set up, desktop VPN clients for common VPN protocols. Our study covers three common desktop operating systems: Windows, macOS and Ubuntu. The study reveals various vulnerabilities in the configurations of VPN clients, which allow attackers to strip off traffic encryption or to bypass server authentication. By exploiting these vulnerabilities, attackers can intercept network traffic to and from the victim. Some of the vulnerabilities also allow the attacker to steal user credentials for to authenticating the VPN gateway. To the best of our knowledge, the vulnerabilities that we present in this paper have not been discussed before in research literature. While each of the vulnerabilities alone might seem like a trivial mistake, together they indicate a serious lack of security-awareness across the commercial VPN industry, and we feel that it is the responsibility of the research community to raise the issue.

We have disclosed all of the vulnerabilities to the responsible parties, and we include the responses that we received so far in the paper. We also provide guidelines on fixing the vulnerabilities. Through this work, we hope to raise awareness among the commercial VPN providers about common configuration mistakes and how they can be avoided.

\boldtitle{Paper organization. }
Section~\ref{sec:background} provides background information on commercial VPN services. Section~\ref{sec:study} gives an overview of our study, and Section~\ref{sec:vulnerabilities} covers the vulnerabilities that were found in the client configurations of popular commercial VPN services. Mitigation solutions are covered in Section~\ref{sec:solutions}. Section~\ref{sec:discussion} discusses the results. Finally, Section~\ref{sec:related} summarizes related work, and Section~\ref{sec:conclusion} concludes the paper.

%!TEX root = ../main.tex

\section{Background}
\label{sec:background}
This section gives an overview of commercial VPN services and their client software.

\subsection{Overview of commercial VPN services}
The focus of this paper is on commercial VPNs, whose usage differs from that of corporate VPNs. The corporate VPNs (also known as enterprise or business VPNs) are set up by organizations to allow their employees to remotely access resources, such as web and file servers, on the corporate intranet. In such networks, system administrators usually pre-configure employee devices with the required client software.

The commercial VPNs (also known as personal or consumer VPNs), on the other hand, are subscription-based services available to regular Internet users. They allow users to tunnel their Internet traffic via the service provider's gateway server somewhere in the cloud.
The commercial VPNs are typically used for personal purposes, such as accessing geoblocked or country-specific media contents and for securing sensitive online activities while on public Wi-Fi networks. They are also used to avoid censorship and surveillance by local governments and access-network operators. However, the users of commercial VPNs route their network traffic through the VPN provider's gateway, which means that they have to trust the VPN provider as well as the country where the gateway server is located. To overcome this trust issue, the commercial VPN providers often promise to keep no logs of the customer activities and allow users to choose VPN servers in countries with strong personal privacy laws (e.g.~Switzerland and Iceland).

\subsection{Commercial VPN client software}
\label{sec:vpn_app}

Most commercial VPN providers have a \textit{native client application}, which sets up the VPN connection for the user. These client applications are usually available for the Windows and macOS operating systems. To use a native client application, users must first enter their VPN user-login credentials into the client application. The application pulls configuration data, such as VPN server addresses, roots of trust for the authentication, and VPN-client credentials (which are not necessarily the same as the user-login credentials), from the VPN provider's server. The client application then configures the VPN tunnel for later use. The client application typically allows the user to choose from many different VPN protocols for the tunnel implementation, either to circumvent firewalls or to match user preferences.

The native client applications rely on the \textit{operating system's built-in VPN client} functionality whenever it exists. They usually use the routing and remote access service (RRAS)~\cite{rras} on Windows and NEVPNManager APIs~\cite{nevpnmanager} on macOS to programmatically create and manage VPN connections. Windows is bundled with implementations of the PPTP, L2TP/IPsec, SSTP and IKEv2 protocols, while macOS comes with L2TP/IPsec, IKEv2 and Cisco IPsec client functionality. For the protocols that have no built-in support in the OS, the commercial VPN providers include \textit{third-party client binaries} in their native applications. Table~\ref{tab:vpn_clients} shows the preferred or recommended VPN clients for each of the studied VPN protocols on the three OSs we consider in this study.

The process of opening a VPN connection is similar regardless of the protocol and the choice between a built-in and third-party client. First, the client establishes a connection to the specified server with the selected protocol. The client and the server then authenticate each other in the selected protocol with the previously configured credentials and roots of trust. After successful authentication, the client and server negotiate various parameters for the VPN connection, such as the encryption scheme and the DNS servers. When the negotiation is completed, the client computer's routing table or firewall rules are configured to tunnel all network traffic through the VPN connection.

The VPN providers do not always have native client applications for all operating systems, such as Linux. Some users may also prefer not to install the provided application. For these users, the VPN providers give instructions on their websites for configuring the OS's built-in VPN client. They may also give advice on installing and configuring third-party clients to use with their service.

%!TEX root = ../main.tex

\section{Study of Commercial VPN Services}
\label{sec:study}

This section gives an overview of our study of commercial VPN services. We first describe the adversary model. We then explain the methodology for systematically finding vulnerabilities in VPN client configuration, covering a large part of the consumer VPN market. The discovered vulnerabilities will be described in detail in the next section.

%!TEX root = ../main.tex

\subsection{Adversary model}
\label{sec:adversary_model}

The object of our study is the way commercial VPN services make use of the common VPN protocols and tunnels. We consider two types of attackers: \textit{network attacker} and \textit{local attacker}. The former is the standard model for network security, while the latter extends the attacker model to non-privileged processes running on the same computer as the VPN client software.

\boldtitle{Network attacker.}
We consider an active network attacker who can intercept and modify network traffic originating from and destined to the user's machine. The attacker could, for example, be a rogue hotspot operator at a hotel or airport, or a compromised core-network operator.

\boldtitle{Local attacker.}
The VPN client software on the user's computer often comprises multiple components that are connected to each other with inter-process communication (IPC). For example, the GUI component may use an IPC channel to sent the VPN configuration to a third-party client binary. It has been recently shown that misconfigured IPC may be vulnerable to attacks by \textit{non-privileged} processes of other users, including guest users, who have access to the same computer (so-called Man-in-the-Machine attacks~\cite{mitma}). These unprivileged attackers could exploit the IPC channels of VPN client applications to steal sensitive information or to modify the VPN connection settings. We included this new type of attackers to the study because the vulnerability of VPN clients to it is currently not well understood and the attacks are different from those on the network. (Note that we do not consider malware that is running with the victim user's privileges or as administrator. While these threats can be serious, the current desktop OSs are not expected to offer protection against them.)

In both the network and local attacks, the ultimate goal of the attackers is to bypass the protection mechanisms of the VPN connection so that they can steal sensitive data sent or received through it.

%!TEX root = ../main.tex

\subsection{Methodology}
\label{sec:methodology}

Given the large number of commercial VPN services that exist today, it is not possible to study all of them. Therefore, we selected 30 of them based on popularity and advertised features (refer to Table~\ref{tab:summary_table}). As a rough estimate of popularity, we searched for ``best VPN services'' on Google and counted how many times each service was mentioned in the resulting pages. The idea was to identify the services that normal users would be most likely to choose. Among the popular commercial VPN services, we prioritized those that support a higher number of VPN protocols.

We focused on the standardized and most commonly supported VPN protocols: \emph{PPTP}, \emph{L2TP/IPsec}, \emph{IKEv2}, \emph{Cisco IPsec}, \emph{SSTP}, \textit{OpenVPN}, and \emph{SoftEther VPN}. We omitted from the study some proprietary protocols (e.g.~OpenWeb and StealthVPN by Astrill~\cite{astrillprotocols}) or in-progress designs (e.g.~WireGuard~\cite{wireguard}) that are not widely deployed in the regular Internet user community.

In addition to looking for new vulnerabilities, we also checked whether the selected commercial VPN providers use publicly known pre-shared keys for L2TP/IPsec (see Section~\ref{sec:intro}). The reason for investigating this known vulnerability is that it severely undermines the security of the users and there was no evidence that VPN service providers had taken the issue seriously.

 We analyzed the selected VPN services with a semi-manual two-step process: (1) \emph{configuration analysis} and (2) \emph{experimental verification}.

\boldtitle{Configuration analysis. }
In each of the commercial VPN client applications, we looked at the way the application creates and configures the VPN connection. When the VPN service providers recommended a built-in client in the OS or a third-party client, we scrutinized the provided configuration instructions and unchanged default settings. In both cases, we looked for potential misconfigurations and architectural mistakes that might compromise the security of the resulting VPN connection. We did not try to find flaws in the cryptographic protocols themselves or code-level implementation errors.

\boldtitle{Experimental verification. }
When we found a potential client-configuration issue, we verified it by implementing an exploit with a set of semi-automated tools built for the purpose. Depending on the type of the attack, the verification was done as follows.

For network attacks, we first created a fake VPN server to intercept connections from the client to the gateway server.
We then routed the VPN client's traffic to the fake server as follows. When testing a commercial VPN client application, we edited \texttt{/etc/hosts} to map the true VPN server's domain name to the fake server's IP address. When testing the instructions for configuring a built-in or third-party VPN client, we simply followed the instruction but gave the fake server's IP address as the gateway address. These methods sufficiently emulate the behavior of a network attacker that intercepts the connections on an untrusted access or core network. Finally, if the VPN client successfully connected to the fake server without dropping the connection or alerting the user, we concluded that the client is vulnerable to the attack currently under test.

We used Poptop~\cite{poptop} to build the fake PPTP server, OpenVPN software~\cite{openvpnsoftware} for the fake OpenVPN server, StrongSwan~\cite{strongswan} for the fake IKEv2 and Cisco IPsec servers, and SoftEtherVPN software~\cite{softethervpnsoftware} for the fake SoftEther VPN, SSTP, and L2TP/IPsec servers (the SoftEtherVPN software supports multiple VPN protocols).

On the other hand, for local attacks, we created two user accounts on a test machine: one acted as the honest user and the other as the attacker. The attacker here is a standard user with no administrative privileges (a guest account can be equally used). We wrote a script to exploit the potential vulnerability in the inter-process communication, executed it in the attacker's login session, and checked whether it succeeded in exploiting the vulnerability in the VPN client application that was running in the honest user's login session.

%!TEX root = ../main.tex

\section{Study results}
\label{sec:vulnerabilities}

This section describes the vulnerabilities that we found in the client configuration of the commercial VPN services selected for this study. In each subsection, we present a brief overview of the VPN protocol, followed by the vulnerabilities related to it. Table~\ref{tab:summary_table} summarizes our findings.

\subsection{Point-to-Point Tunneling Protocol}
\label{sec:pptp}

Point-to-Point Tunneling Protocol (PPTP)~\cite{hamzeh1999point} was created by Microsoft, and it is one of the oldest VPN protocols. It has well-known weaknesses~\cite{marlinspike2012divide,schneier1999cryptanalysis,horst2016breaking,mudge1998cryptanalysis} and is no longer considered secure. Nevertheless, the protocol remains widely deployed and used because many firewalls do not block it. Our goal is to analyze the PPTP configurations of commercial VPN services to see whether they have additional weaknesses that could further compromise the users' security and privacy.

\begin{figure}[ht]
  \centering
  \includegraphics[width=0.85\columnwidth]{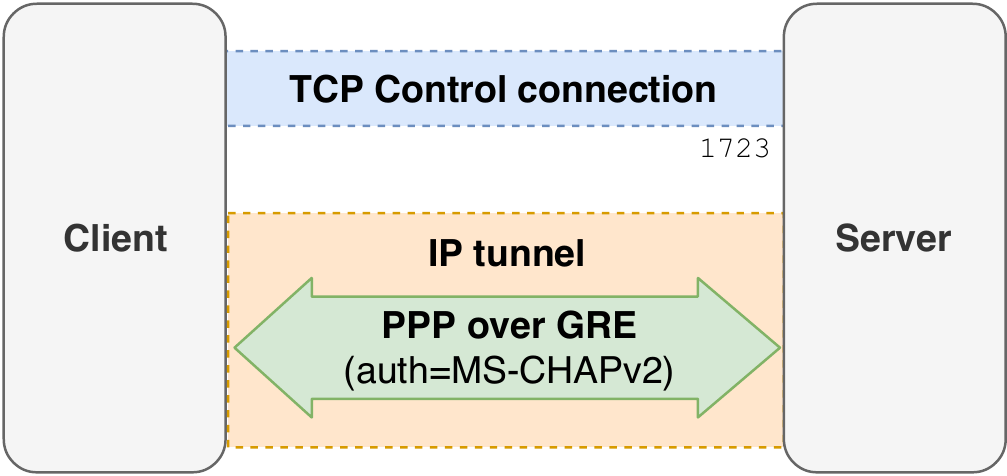}
  \caption{Typical PPTP connection}
  \label{fig:pptp}
\end{figure}

There are two parallel parts im a PPTP connection: a \textit{TCP control connection} and an \textit{IP tunnel}, as illustrated in Figure~\ref{fig:pptp}. To instantiate a PPTP tunnel, the client first establishes the control connection to port 1723 of the PPTP server. The control connection is then used to initiate and manage an IP tunnel between the client and the server, and a Point-to-Point Protocol (PPP)~\cite{simpson1994ppp} session inside the IP tunnel. PPTP uses Generic Routing Encapsulation (GRE)~\cite{farinacci2000generic} to encapsulate PPP packets. The encapsulation also supports flow control and congestion control on the tunnel. The GRE header in PPTP differs slightly from the GRE specification by having an acknowledgment number field. This field is used to determine whether a particular GRE packet or set of packets has arrived at the other endpoint of the tunnel~\cite{hamzeh1999point}.

The PPP session goes through the following three phases:

\begin{enumerate}
  \item \textit{Authentication}: The communication endpoints must first authenticate each other. PPP supports various authentication methods; however, most commercial VPN services implement only MS-CHAPv2.
  \item \textit{Negotiation}: If the authentication is successful, the client and the server negotiate parameters such as the encryption scheme and the DNS servers for the client. The only encryption scheme that PPP supports is Microsoft Point-to-Point Encryption (MPPE)~\cite{pall2001microsoft}. A well-configured PPTP server usually enforces MPPE with 128-bit keys on its connections. The encryption key is derived from the authentication in the previous phase. It is important to note that all packets until completion of the negotiation phase are transmitted in plain text.
  \item \textit{Data exchange}: Finally, the client starts communicating network traffic with the server by encrypting it as per the negotiated encryption scheme.
\end{enumerate}

\boldtitle{Optional encryption.}
Windows by default does not enforce encryption on any VPN connection. We found that many of the commercial VPN services in our study do not instruct their users to change this setting while configuring PPTP with the built-in client on Windows. A network attacker can take advantage of this behavior to perform server impersonation as follows. First, the attacker acts as a man-in-the-middle to forward traffic between the client and the honest server until the authentication phase is finished. The attacker then switches to performing server impersonation and negotiates with the client not to encrypt the data exchange. The client agrees to this because it is not mandatory to use encryption. As the result, the attacker obtains all traffic of the victim just as if no VPN was used.

% ====================================================================================

\subsection{SSTP}
\label{sec:sstp}

\begin{figure}[ht]
  \centering
  \includegraphics[width=0.85\columnwidth]{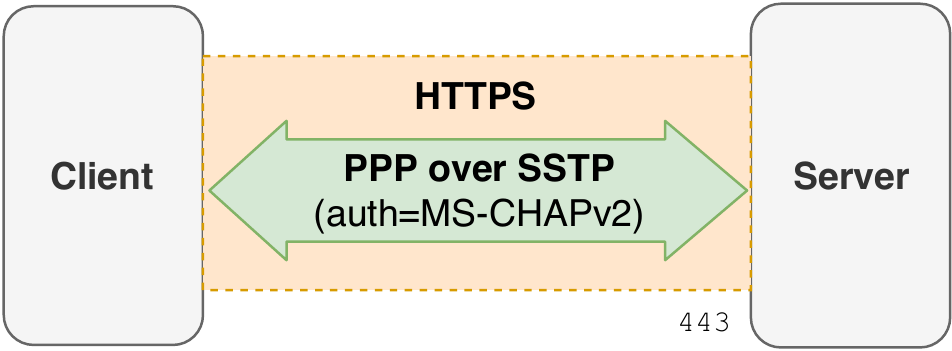}
  \caption{Typical SSTP connection}
  \label{fig:sstp}
\end{figure}

Secure socket tunneling protocol (SSTP)~\cite{sstp} is another VPN protocol created by Microsoft. Figure~\ref{fig:sstp} shows an overview of a typical SSTP connection. It utilizes PPP to transport network traffic, but instead of using GRE as PPTP does, it encapsulates the PPP packets in HTTPS.

An SSTP connection is established as follows:

\begin{enumerate}
  \item First, the client opens an HTTPS connection to the server. The client authenticates the server by verifying the server's TLS certificate as in any HTTPS connection.

  \item If the TLS authentication succeeds, the client begins SSTP negotiation by sending a \texttt{Connect-Request} message to the server. The server replies with a \texttt{Connect-Acknowledgement} message that contains a nonce to be used later.

  \item Both sides perform the PPP authentication, deriving a session key for MPPE. Like in PPTP, MS-CHAPv2 is usually used for this. However, in this case, MS-CHAPv2 is protected against active and passive attacks by the HTTPS encryption.

  \item \label{step:sstp_connected} The client sends a \texttt{Call-Connected} message that contains the nonce received from the server, a hash of the server certificate from the HTTPS handshake, and a message authentication code (MAC) that is computed over the message with the MPPE key derived during the PPP authentication. This message cryptographically binds the PPP session to the server identity from the outer TLS authentication.

  \item The endpoints perform the PPP negotiation and then start to exchange network traffic. SSTP does not use MPPE encryption. Instead, it relies entirely on HTTPS for the secure delivery of its messages.
\end{enumerate}

Windows has a built-in SSTP client. On macOS and Ubuntu, VPN services usually suggest the user to install \textit{EasySSTP} and \textit{sstp-client}, respectively~\cite{purevpnsstpubuntu,torguardsstpubuntu,strongvpnsstpubuntu,unblockvpnsstpubuntu}. EasySSTP~\cite{easysstp}, however, has not received any updates since 2013, and it no longer works on the latest version of macOS (10.14) at the time of writing this paper.

\boldtitle{Ignored certificate verification failures.}
We tested the SSTP client on Windows 10 and the latest sstp-client (v1.0.11) to see whether they perform the mutual authentication properly. While the Windows client does this correctly,  \texttt{sstp-client} on Ubuntu does not consider whether the server's certificate is trusted. By inspecting the source code of \texttt{sstp-client}~\cite{sstpclient}, we found the reason for this unexpected behavior: the \texttt{sstp-client} is integrated into Ubuntu network connection manager, which allows the user to configure whether the connection should be terminated when the certificate verification fails, but \texttt{sstp-client} ignores certificate verification errors regardless of this setting.

Ignoring the certificate verification failure allows the network attacker to perform a server impersonation attack as follows. First, the fake server presents a self-signed TLS certificate to the honest client. The fake server then connects to the honest server pretending to be the client. The attacker forwards traffic between the honest client and the honest server until the PPP authentication is completed and the attacker sees the \texttt{Call-Connected} message (Step~\ref{step:sstp_connected} above). The attacker then stops forwarding traffic to the honest server and finishes the PPP negotiation by itself. When the SSTP connection is successfully established, the attacker's fake server can act as the VPN gateway and obtain all the victim's traffic.

\subsection{IKEv2}
\label{sec:ikev2}

IKEv2 VPN is a more modern VPN protocol based on IPsec. It uses IKEv2~\cite{kaufman2014internet} for authentication as well as establishing and maintaining security associations. One improvement of IKEv2 over IKEv1 is that the new protocol allows each endpoint to use a different authentication method. IKEv2 also supports EAP ~\cite{aboba2004extensible}, extending the selection of available authentication methods.

\begin{figure}[ht]
  \centering
  \includegraphics[width=0.9\columnwidth]{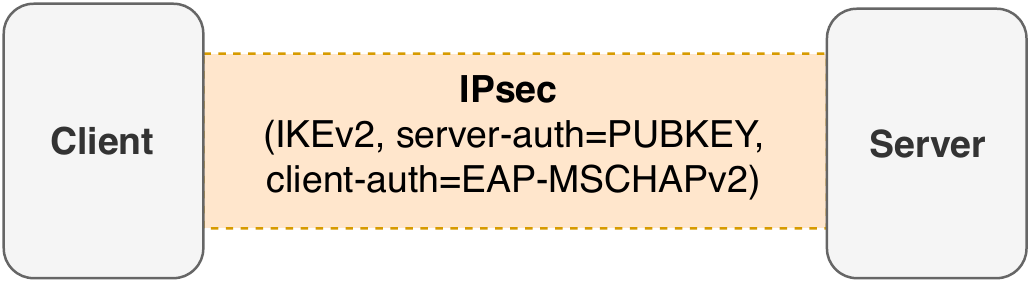}
  \caption{Typical IKEv2 VPN connection}
  \label{fig:ikev2}
\end{figure}

Figure~\ref{fig:ikev2} shows a typical setup by VPN services. The server authenticates itself to the client with a certificate while the client authenticates to the server with EAP-MSCHAPv2~\cite{hurst2007microsoft}, which is basically MS-CHAPv2 encapsulated in the EAP protocol. The IKEv2 connection establishment works as follows:

\begin{enumerate}
  \item \textit{Initial exchange}: The client and server negotiate security parameters, such as cryptographic algorithms, and exchange nonces and Diffie-Hellman (DH) values. After that, each party computes the shared session keys, which will be used for protecting all the following messages. These values will also be used for constructing the first security association (SA).

  \item \textit{Server authentication}: The server authenticates itself to the client with its certificate (or certificate chain) and a signature on the SA data.

  \item \textit{Client authentication with EAP: } The client is then authenticated with the EAP-MSCHAPv2 protocol. After completion of EAP-MSCHAPv2 protocol, the client and server exchange MACs to bind the EAP authentication to the created SA. The MACs are calculated over the SA data with the session key produced by the MS-CHAPv2 protocol.
\end{enumerate}

Both Windows and macOS have built-in support for IKEv2. On Linux systems such as Ubuntu, commercial VPN services usually instruct their users to install StrongSwan~\cite{strongswan}, an open-source IPsec implementation, for the client.

\boldtitle{Unspecified server name.}
To establish an IKEv2 connection with StrongSwan on Ubuntu, the user has to create a profile for the connection in \texttt{/etc/ipsec.conf}. Several commercial VPN providers in our study instruct their users to create the profile as follows (only important parts are shown).

\lstset{xleftmargin=0.5cm}
\begin{lstlisting}
leftauth=eap-mschapv2
...
right=<server-address>
rightauth=pubkey
rightid=%any
...
\end{lstlisting}

Left and right indicate the client and the server, respectively. With such a profile, the server uses a public key for authentication while the client uses EAP-MSCHAPv2. The problem is with the \texttt{rightid} setting, which tells how the server should be identified in the authentication. Since it is set to \texttt{\%any}, the client will accept any certified server regardless of its identity. Consequently, the network attacker can pick any domain that it owns and purchase a certificate from a widely trusted CA. The attacker can then impersonate the server in the server authentication step because the VPN client does not check the name in the certificate.

However, MS-CHAPv2 actually provides mutual authentication with the user password. The binding of this authentication to the SA prevents the attacker from completing the protocol without knowing the user password. Thus, the misconfiguration effectively reduces the security of IKEv2 to that of MS-CHAPv2, which unfortunately is equal to the strength of a single DES encryption and also vulnerable to password cracking \cite{schneier1999cryptanalysis,marlinspike2012divide}. The resulting security is significantly weaker than expected because usually these weaknesses of MS-CHAPv2 are masked when it is used inside a server-authenticated tunnel.

% ====================================================================================

\subsection{OpenVPN}
\label{sec:openvpn}

OpenVPN~\cite{openvpn} appears to be the most widely supported protocol by commercial VPN services. It uses TLS as the underlying authentication and key exchange protocol. Commercial VPN services deploy OpenVPN in the client-server mode. In this mode, the server authenticates itself to the client with an X.509 certificate signed by a CA that the client trusts while the client proves its identity to the server with a username and password.

Despite the wide range of configuration options that OpenVPN supports, we did not find any broken configuration examples that would allow the network attacker to compromise the OpenVPN connection. This is probably because OpenVPN has very detailed documentation and configuration guidelines. We found, however, that local attackers, i.e.~non-privileged local users and processes (see Section~\ref{sec:adversary_model}), can steal the username and password that are used for authenticating the client.

Most OSs do not provide native support for OpenVPN. Thus, commercial VPN services have to include the open-source \texttt{openvpn} client binary~\cite{openvpnsoftware} in their client software to support the protocol. The binary is run as a daemon, which creates the VPN connection to the server. For this, it needs configuration information including the server address, server name, and trusted CA certificate. There are two ways of delivering the configuration information to the \texttt{openvpn} daemon: the client GUI application can either write the configuration to a file or pass it to the daemon as command-line parameters. Additionally, the daemon will need the client's VPN username and password, which can be specified either in the configuration file or via the management interface described below.

The OpenVPN daemon supports a management interface~\cite{openvpnmanagement}, which allows administrative control via a TCP connection. By default, it only accepts connections on the \texttt{localhost} interface, and it can be configured to require password authentication of the administrator. The advantage of the management interface is that the user can avoid saving the client credentials into the configuration file on the disk. The commercial VPN application first starts the OpenVPN daemon with all the necessary configuration options, except the client credentials, and puts it on hold with the \texttt{management-hold} option. The application then connects to the management interface of the daemon, gives it the username and password, and finally releases the connection from the hold state.

\boldtitle{Credential leakage.}
Some commercial VPN client applications are careless when passing the username and password to the OpenVPN daemon. Specifically, the VPN applications store the credentials in configuration files that are readable to all users on the client computer. On Windows, the configurations are usually stored in the \texttt{ProgramData} or \texttt{Program File} folders, and the vulnerable software does not set the access-control list to prevent unauthorized users from accessing it. Therefore, the local attacker (i.e.\ any un-privileged process running on the same computer) can capture this sensitive information. Some of these services remove the credentials from the file after the connection has been established, but this still leaves a window of a few seconds to capture the information.

\boldtitle{Lack of admin password.}
Another issue with most of the commercial VPN services in our study is that they do not enable password protection on the management interface. We do not know of a way to exploit this weakness, except denial-of-service attacks to annoy the victim user. However, we would not be surprised if the lack of authentication is later found to enable exploits by local attackers.

% ====================================================================================

\subsection{SoftEther VPN}
\label{sec:softether}

SoftEther VPN~\cite{softethervpn} is another VPN protocol with an open-source implementation that tunnels Ethernet frames over HTTPS. Similar to OpenVPN in the client-server mode, the SoftEther VPN server proves its identity to the client with a TLS certificate while the client has a username and password for its authentication.

The SoftEther VPN connection establishment is implemented in two binaries: \texttt{vpncmd}, the command-line administrative tool, and \texttt{vpnclient}, its command execution worker. The \texttt{vpnclient} worker process runs a TCP server on port 5555 for receiving administrative commands. By default, the TCP server accepts connections only from the \texttt{localhost} interface. It can be configured to require password authentication.

By issuing commands to the \texttt{vpnclient} TCP server, \texttt{vpncmd} can perform administrative operations such as creating a new VPN profile or editing an existing one, or starting and stopping a VPN connection with an existing profile. SoftEther VPN profiles have various configurable options, including whether the client should check the server's TLS certificate or not.

\boldtitle{No server verification.}
Hide.me supports SoftEther VPN on its GUI. However, the~\texttt{CheckServerCert} parameter is set to \texttt{false} in its client configuration. Consequently, the client does not verify the server's certificate. This allows the network attacker to perform a MitM attack on its connections. The attacker can thus obtain the victim's credentials and network traffic.

Some other commercial VPN services support SoftEther VPN by providing instructions on how to connect to their servers with the SoftEther VPN GUI application~\cite{softethervpnsoftware}. When creating a new SoftEther VPN connection, the GUI allows the user to choose whether the client verifies the server's certificate. The default setting is \texttt{False}, and the VPN services do not tell their users to change the setting. Thus, they are vulnerable to the same attacks as Hide.me.

\boldtitle{Wrong VPN server.}
We found another problem with Hide.me, which is that its GUI does not require password authentication on the management interface of the \texttt{vpnclient} process. This allows the local attacker to connect to the interface and, for example, launch a new VPN connection that routes all network traffic of the victim to a malicious server under the attacker's control. This is a typical case of the man-in-the-machine attack on IPC \cite{mitma}.

% ====================================================================================
%!TEX root = ../main.tex

% center p-type cell for tables
\newcolumntype{x}[1]{>{\centering\arraybackslash\hspace{0pt}}p{#1}}
\newcommand{\colw}{1.2cm}
\newcommand{\headbox}[1]{\rotatebox{60}{\parbox{2.9cm}{#1}}}

\begin{table*}[ht!]
    \centering
    \def\arraystretch{1.3}
    \begin{tabularx}{\textwidth}{l| x{\colw} x{\colw}  x{\colw} x{\colw} x{\colw} x{\colw} x{\colw} x{\colw} x{\colw}  }
        & \headbox{PPTP: \\ Optional encryption}
        & \headbox{SSTP: \\ Ignored certificate \\verification failure}
        & \headbox{IKEv2: \\ Improper server \\verification}
	    & \headbox{OpenVPN: \\ Credential leakage}
        & \headbox{SoftEther: \\ No server verification}
        & \headbox{SoftEther: \\ Wrong VPN server}
        & \headbox{L2TP/IPsec: \\ Known pre-shared key}
        & \headbox{Cisco IPsec: \\ Known pre-shared key}
        & \headbox{Fallback \\ to weak protocol}
        \\

        \hline Operat typeing systems
            & W % PPTP enc
            & U % SSTP
            & U % IKEv2
            & W % OpenVPN
            & W, M % SoftEther server verification
            & W, M % SoftEther evil vpn
            & W, M, U % L2TP
            & W, M, U % Cisco
            & W, M % Auto
            \\

        \hline Attacker type
            & Network % PPTP enc
            & Network % SSTP
            & Network % IKEv2
            & Local % OpenVPN
            & Network % SoftEther server verification
            & Local % SoftEther evil vpn
            & Network % L2TP
            & Network % Cisco
            & Network % Auto
            \\

        \hline

        \hline Astrill
            & \notvuln % PPTP enc
            & \na % SSTP
            & \na % IKEv2
            & \notvuln % OpenVPN
            & \na % SoftEther server verification
            & \na % SoftEther evil vpn
            & \vuln % L2TP
            & \vuln % Cisco
            & \na % Auto
            \\

        \hline BoxPN
            & \vuln % PPTP enc
            & \na % SSTP
            & \na % IKEv2
            & \vuln % OpenVPN
            & \na % SoftEther server verification
            & \na % SoftEther evil vpn
            & \vuln % L2TP
            & \vuln % Cisco
            & \na % Auto
            \\

        \hline CactusVPN
            & \vuln % PPTP enc
            & \na % SSTP
            & \na % IKEv2
            & \notvuln % OpenVPN
            & \vuln % SoftEther server verification
            & \na % SoftEther evil vpn
            & \vuln % L2TP
            & \na % Cisco
            & \na % Auto
            \\

        \hline CyberGhost
            & \notvuln % PPTP enc
            & \na % SSTP
            & \na % IKEv2
            & \notvuln % OpenVPN
            & \na % SoftEther server verification
            & \na % SoftEther evil vpn
            & \vuln % L2TP
            & \vuln % Cisco
            & \notvuln % Auto
            \\

        \hline ExpressVPN
            & \na % PPTP enc
            & \na % SSTP
            & \na% IKEv2
            & \vuln % OpenVPN
            & \na % SoftEther server verification
            & \na % SoftEther evil vpn
            & \vuln % L2TP
            & \na % Cisco
            & \vuln % Auto
            \\

        \hline FastestVPN
            & \notvuln % PPTP enc
            & \na % SSTP
            & \na % IKEv2
            & \vuln % OpenVPN
            & \na % SoftEther server verification
            & \na % SoftEther evil vpn
            & \vuln % L2TP
            & \vuln % Cisco
            & \na % Auto
            \\

        \hline FrootVPN
            & \vuln % PPTP enc
            & \na % SSTP
            & \na % IKEv2
            & \vuln % OpenVPN
            & \na % SoftEther server verification
            & \na % SoftEther evil vpn
            & \vuln % L2TP
            & \vuln % Cisco
            & \na % Auto
            \\

        \hline GooseVPN
            & \na % PPTP enc
            & \na % SSTP
            & \na % IKEv2
            & \vuln % OpenVPN
            & \na % SoftEther server verification
            & \na % SoftEther evil vpn
            & \vuln % L2TP
            & \vuln % Cisco
            & \vuln % Auto
            \\

        \hline Hide.me
            & \notvuln % PPTP enc
            & \na % SSTP
            & \vuln % IKEv2
            & \notvuln % OpenVPN
            & \vuln % SoftEther server verification
            & \vuln % SoftEther evil vpn
            & \vuln % L2TP
            & \na % Cisco
            & \notvuln% Auto
            \\

        \hline HideMyAss
            & \vuln % PPTP enc
            & \na % SSTP
            & \na % IKEv2
            & \vuln % OpenVPN
            & \na % SoftEther server verification
            & \na % SoftEther evil vpn
            & \vuln % L2TP
            & \vuln % Cisco
            & \na % Auto
            \\

        \hline ibVPN
            & \vuln % PPTP enc
            & \na % SSTP
            & \na % IKEv2
            & \notvuln % OpenVPN
            & \vuln % SoftEther server verification
            & \na % SoftEther evil vpn
            & \vuln % L2TP
            & \vuln % Cisco
            & \notvuln % Auto
            \\

        \hline IPVanish
            & \vuln % PPTP enc
            & \na % SSTP
            & \na % IKEv2
            & \vuln % OpenVPN
            & \na % SoftEther server verification
            & \na % SoftEther evil vpn
            & \vuln % L2TP
            & \vuln % Cisco
            & \na % Auto
            \\

        \hline IVPN
            & \na % PPTP enc
            & \na % SSTP
            & \notvuln % IKEv2
            & \notvuln % OpenVPN
            & \na % SoftEther server verification
            & \na % SoftEther evil vpn
            & \na % L2TP
            & \na % Cisco
            & \na % Auto
            \\

        \hline LimeVPN
            & \vuln % PPTP enc
            & \na % SSTP
            & \na % IKEv2
            & \na % OpenVPN
            & \vuln % SoftEther server verification
            & \na % SoftEther evil vpn
            & \vuln % L2TP
            & \na % Cisco
            & \na % Auto
            \\

        \hline NordVPN
            & \na % PPTP enc
            & \na % SSTP
            & \vuln % IKEv2
            & \notvuln % OpenVPN
            & \na % SoftEther server verification
            & \na % SoftEther evil vpn
            & \na % L2TP
            & \na % Cisco
            & \na % Auto
            \\

        \hline OverplayVPN
            & \notvuln % PPTP enc
            & \na % SSTP
            & \na % IKEv2
            & \notvuln% OpenVPN
            & \na % SoftEther server verification
            & \na % SoftEther evil vpn
            & \vuln % L2TP
            & \na % Cisco
            & \na % Auto
            \\

        \hline Perfect-Privacy
            & \na % PPTP enc
            & \na % SSTP
            & \na % IKEv2
            & \notvuln % OpenVPN
            & \na % SoftEther server verification
            & \na % SoftEther evil vpn
            & \vuln % L2TP
            & \vuln % Cisco
            & \notvuln % Auto
            \\

        \hline PersonalVPN
            & \vuln % PPTP enc
            & \na % SSTP
            & \notvuln % IKEv2
            & \na % OpenVPN
            & \na % SoftEther server verification
            & \na % SoftEther evil vpn
            & \vuln % L2TP
            & \vuln % Cisco
            & \na % Auto
            \\

        \hline Private Internet Access
            & \notvuln % PPTP enc
            & \na % SSTP
            & \na % IKEv2
            & \notvuln % OpenVPN
            & \na % SoftEther server verification
            & \na % SoftEther evil vpn
            & \vuln % L2TP
            & \na % Cisco
            & \na % Auto
            \\

        \hline PrivateVPN
            & \notvuln % PPTP enc
            & \na % SSTP
            & \na % IKEv2
            & \notvuln % OpenVPN
            & \na % SoftEther server verification
            & \na % SoftEther evil vpn
            & \vuln % L2TP
            & \vuln % Cisco
            & \na % Auto
            \\

        \hline ProXPN
            & \na % PPTP enc
            & \na % SSTP
            & \na % IKEv2
            & \notvuln % OpenVPN
            & \na % SoftEther server verification
            & \na % SoftEther evil vpn
            & \na % L2TP
            & \na % Cisco
            & \vuln % Auto
            \\

        \hline PureVPN
            & \vuln % PPTP enc
            & \vuln % SSTP
            & \na % IKEv2
            & \vuln % OpenVPN
            & \vuln % SoftEther server verification
            & \na % SoftEther evil vpn
            & \vuln % L2TP
            & \na % Cisco
            & \notvuln % Auto
            \\

        \hline RocketVPN
            & \na % PPTP enc
            & \na % SSTP
            & \na % IKEv2
            & \na % OpenVPN
            & \vuln % SoftEther server verification
            & \na % SoftEther evil vpn
            & \vuln % L2TP
            & \na % Cisco
            & \na % Auto
            \\

        \hline SaferVPN
            & \vuln % PPTP enc
            & \na % SSTP
            & \na % IKEv2
            & \notvuln % OpenVPN
            & \na % SoftEther server verification
            & \na % SoftEther evil vpn
            & \vuln % L2TP
            & \na % Cisco
            & \vuln % Auto
            \\

        \hline StrongVPN
            & \vuln % PPTP enc
            & \vuln % SSTP
            & \vuln % IKEv2
            & \vuln % OpenVPN
            & \na % SoftEther server verification
            & \na % SoftEther evil vpn
            & \vuln % L2TP
            & \vuln % Cisco
            & \na % Auto
            \\

        \hline TorGuard
            & \vuln % PPTP enc
            & \vuln % SSTP
            & \na % IKEv2
            & \notvuln % OpenVPN
            & \na % SoftEther server verification
            & \na % SoftEther evil vpn
            & \vuln % L2TP
            & \vuln % Cisco
            & \na % Auto
            \\

        \hline Trust.Zone
            & \na % PPTP enc
            & \na % SSTP
            & \na % IKEv2
            & \na % OpenVPN
            & \vuln % SoftEther server verification
            & \na % SoftEther evil vpn
            & \vuln % L2TP
            & \na % Cisco
            & \na % Auto
            \\

        \hline UnblockVPN
            & \vuln % PPTP enc
            & \vuln % SSTP
            & \na % IKEv2
            & \na % OpenVPN
            & \na % SoftEther server verification
            & \na % SoftEther evil vpn
            & \vuln % L2TP
            & \na % Cisco
            & \na % Auto
            \\

        \hline VyprVPN
            & \vuln % PPTP enc
            & \na % SSTP
            & \na % IKEv2
            & \notvuln % OpenVPN
            & \na % SoftEther server verification
            & \na % SoftEther evil vpn
            & \notvuln % L2TP
            & \na % Cisco
            & \na % Auto
            \\

        \hline 24VC
            & \na % PPTP enc
            & \na % SSTP
            & \na % IKEv2
            & \na % OpenVPN
            & \vuln % SoftEther server verification
            & \na % SoftEther evil vpn
            & \vuln % L2TP
            & \na % Cisco
            & \na % Auto
            \\

    \end{tabularx}
    \caption{Summary of discovered vulnerabilities (\vuln : vulnerable, \notvuln : not vulnerable, \na: not applicable, \\ W: Windows, M: macOS, U: Ubuntu)}

    \label{tab:summary_table}
\end{table*}

\subsection{L2TP/IPsec}
\label{sec:l2tp}
As mentioned in Section~\ref{sec:intro}, many commercial VPN services were using known pre-shared keys to authenticate the IPsec tunnel in 2016~\cite{l2tpkeys}. Now, several years later, 26 out of the 30 commercial VPN services in our study still have this vulnerability, despite the fact that there are solutions available (see Section~\ref{sec:solutions}).

\begin{figure}[ht]
  \centering
  \includegraphics[width=0.85\columnwidth]{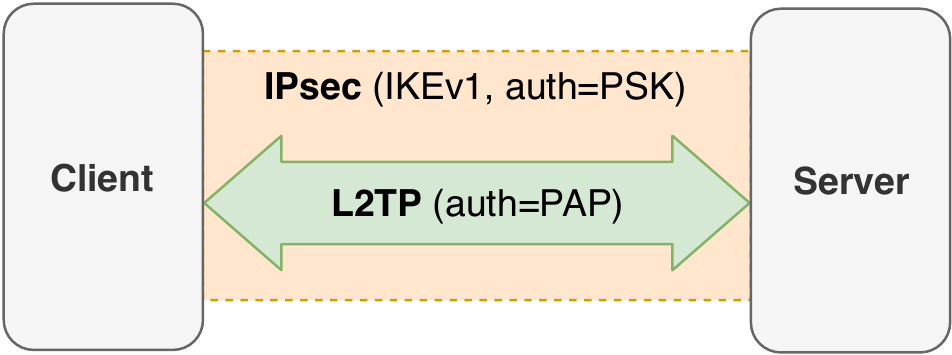}
  \caption{Typical L2TP/IPsec connection}
  \label{fig:l2tp}
\end{figure}

In the L2TP/IPsec VPN connection, L2TP~\cite{townsley1999layer} and IPsec together provide two layers of authentication, as shown in~Figure~\ref{fig:l2tp}: (1) IKEv1 for authenticating the communicating machines and (2) Password Authentication Protocol (PAP) or Challenge-Handshake Authentication Protocol (CHAP)~\cite{lloyd1992ppp} for authenticating the user. However, the confidentiality and integrity of the resulting connection rely entirely on the IPsec tunnel because L2TP does not provide any transport protection. Thus, if the network attacker manages to break the IPsec tunnel, the integrity and confidentiality of the connection are compromised. By using a known pre-shared key to authenticate the IPsec tunnel, commercial VPN providers enable the network attacker to perform a MitM attack on the L2TP/IPsec connection and to capture the victim's network traffic.

% ====================================================================================

\subsection{Cisco IPsec}
\label{sec:cisco}
\begin{figure}[ht]
  \centering
  \includegraphics[width=0.85\columnwidth]{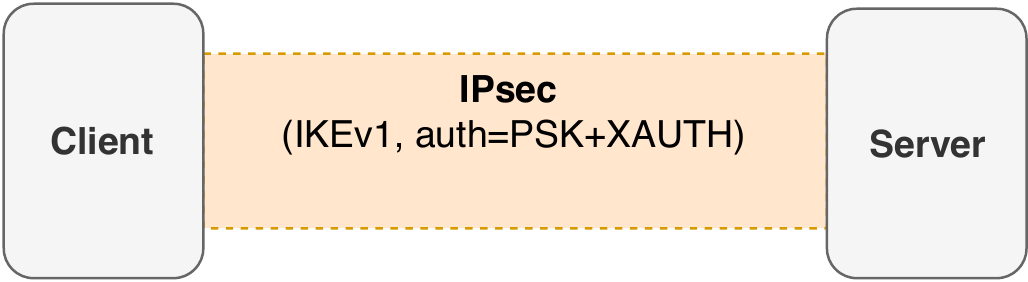}
  \caption{Typical Cisco IPsec connection}
  \label{fig:cisco}
\end{figure}

Cisco IPsec~\cite{ciscoIPsec} is widely used in enterprise VPNs. However, it is also supported by a number of commercial VPN services. Like L2TP/IPsec, the protocol uses IPsec to tunnel traffic between the client and the server, as illustrated in Figure~\ref{fig:cisco}. The main distinguishing feature of Cisco IPsec is that, after the communicating nodes have completed the conventional IKEv1 authentication, an additional phase of Extended Authentication (XAUTH)~\cite{pereira1999extended} is performed to authenticate the user. XAUTH allows various types of user authentication, such as challenge-response and one-time password.

\boldtitle{Known pre-shared keys.}
We found that all of the commercial VPN services that support Cisco IPsec in our study use a pre-shared key to authenticate the IPsec tunnel. Learning from the experience of L2TP/IPsec, it is rather obvious to ask where the endpoints get the pre-shared key. The user interfaces of the commercial VPN clients do not have a way of entering such a key. Perhaps unsurprisingly, we found that the commercial VPN services have fixed pre-shared keys also for Cisco IPsec. This allows the network attacker to perform MitM attacks on these IPsec connections and to obtain all the network traffic.

% ====================================================================================

\subsection{Fallback strategy}
\label{sec:fallback}

As mentioned in Section~\ref{sec:background}, commercial VPN services usually have a list of protocols from which the user can choose on the GUI of the provided application. A special option that several services in our study support is \textit{Automatic}, which means that the application will automatically select the protocol for the user. The way this works is that the application tries different protocols one after another until it succeeds in creating a VPN connection. Users typically choose this option if the firewall in the access network blocks some VPN protocols, or if they do not want to understand the technical intricacies of choosing the right protocol.

\boldtitle{Fallback to weak option.}
It is easy to see that the security of the fallback strategy is equal to that of the weakest option which the application is willing to try. The network attacker can simply block all other connection attempts. We found that some commercial VPN services include L2TP/IPsec with publicly-known pre-shared key as an option in their automatic mode. For example, ExpressVPN attempts to use the following protocols: OpenVPN, SSTP, and L2TP/IPsec. If the network attacker blocks the first two (e.g. by filtering the corresponding ports), it effectively forces the client to use L2TP/IPsec. As the result, the attacker can perform a MitM attack on the VPN connection and obtain the traffic.

Windows also provides similar fallback strategies when RRAS is used to create the VPN connection. RRAS allows the developers to choose the order in which it attempts VPN protocols until the connection is successfully established. This is configured with the \texttt{VpnStrategy} option~\cite{rrasvpnstrategy}. ProXPN, when configuring its IKEv2 client on Windows, instead of setting \texttt{VpnStrategy} to 5 (i.e. attempting IKEv2 only), sets the option to 8, which effectively tells the client to try the following protocols in order: IKEv2, SSTP, PPTP, and L2TP/IPsec. This causes the application to suffer from the same vulnerability in the automatic mode as ExpressVPN.

% ====================================================================================

%!TEX root = ../main.tex

\section{Mitigation solutions}
\label{sec:solutions}
In this section, we discuss potential solutions to the issues presented in Section~\ref{sec:vulnerabilities}.

\boldtitle{PPTP.}
As explained in Section~\ref{sec:pptp}, PPTP encryption is optional in the Windows implementation.  Fixing this issue is straightforward: the VPN service providers simply need to update their instructions to tell Windows users to change the \textit{Data encryption} setting of the PPTP connection adapter from \textit{Optional encryption} to \textit{Maximum strength encryption}. This enforces MPPE encryption with a 128-bit key on the connection. A more sustainable solution would be for Windows to employ strong encryption by default. While such changes to default settings are not always feasible for backward compatibility reasons, \textit{secure by default} is one of the key principles in designing secure systems.

\boldtitle{SSTP.} 
The \texttt{sstp-client} in Ubuntu ignores certificate verification failures. As pointed out in Section~\ref{sec:sstp}, the flaw is in the \texttt{sstp-client} library code. Until it is fixed, commercial VPN services should explicitly instruct their users to not use ~\texttt{sstp-client} and possibly provide an alternative. 
The broader issue here is that modern software development practices create complex dependencies on free and third-party components, for which there may not be guaranteed maintenance. One would expect security-critical services, such as commercial VPNs, to manage their dependencies carefully. 

\boldtitle{IKEv2.}
Setting the \texttt{rightid} value of a StrongSwan's IKEv2 configuration to \texttt{\%any} could be useful, for example, when testing a VPN server. However, the setting should never be used in production. To fix the problem, the commercial VPN providers should give clear instructions to the users to set \texttt{rightid} to the server's domain name or to the Distinguished Name from the server's certificate~~\cite{strongswanIDParsing}.
Alternatively, if the \texttt{right} parameter (i.e. the address of the server) is already set to the server's domain name, \texttt{rightid} does not need to be configured because it equals the value of \texttt{right} by default.

The security failure here arises from the fact that it is easier to get the service to work with insecure wildcard settings than to find out and configure the correct server name. Even if the VPN service providers documented the correct usage, some users would find easier, insecure configuration entries online. Probably the only safe solution in this case is to automate the setup process and to audit the configuration regularly for unsafe changes by the user. 

\boldtitle{OpenVPN.}
The obvious solution to unauthorized users reading the OpenVPN configuration file is to set the access controls on the file, which is world-readable by default. Another solution is to avoid writing the VPN username and password to the file by communicating them over the management interface. Password authentication on the management interface is also a good practice, even though the lack of authentication does not result in known attacks in OpenVPN.

Storing secrets on the local machine is a problem encountered commonly by security-critical software. There is no perfect solution but making use of operating-systems access controls to is a good starting point. The next step might be a secure hardware module for storing client credentials.  

\boldtitle{SoftEther VPN.}
Access to the management interface is far more critical in SoftEther VPN than in OpenVPN because, without authentication of the administrator, it allows any local user or process to control the VPN connections. Thus, the management interface must be protected with proper password authentication. The higher-level issue here is that TCP connections to the localhost are not inherently secure and may require application-level authentication between the client and the server \cite{mitma}. 

To fix the server verification problem that we described, the \texttt{CheckServerCert} parameter of the SoftEther VPN connection must be set to \texttt{true} so that the client verifies the server's certificate during the establishment of the connection. This can be done either by changing the default value of \texttt{CheckServerCert} in SoftEther, or the VPN services can provide explicit instructions to their customers to do the same. This again highlights the importance of safe default values and the danger of allowing easy but insecure settings.

\boldtitle{L2TP/IPsec and Cisco IPsec.}
% \boldtitle{Cisco IPsec.}
The commercial VPN services that support the L2TP/IPsec and Cisco IPsec protocols share the same problem of using known pre-shared key for the IPsec authentication. Before considering solutions, it is worth discussing the reason for the use of fixed keys in these IPsec-based VPN protocols. They both use IKEv1 in the Main Mode, in which the server selects the pre-shared key by the IP address of the client. IKEv1 and the entire IPsec architecture reside in the IP layer, and thus the IP address is the only clue available for them about the client identity. Since the clients of the commercial VPN services practically always have dynamic IP addresses, it is not possible for the VPN gateway to support client-specific pre-shared keys. There have been proprietary proposals for sending a hint about the client identity to the server, but the rather historical IKEv1 protocol and its implementations have not been updated to support such new features in an interoperable way. Thus, the commercial VPN services have fallen back to the insecure practice of sharing the same key between all clients.

A solution is to switch from pre-shared keys to certificate authentication. For this, the commercial VPN services must obtain certificates for their VPN servers from a widely trusted commercial CA. The client certificates, on the other hand, can be provisioned by the VPN service provider itself. Client-side authentication in IKEv1 is not extremely critical anyway because the client user is authenticated separately with username and password in the later phases of the VPN protocols. It would, nevertheless, be a good practice to authenticate both the client device (with a certificate) and the client user (with username and password), so that devices and users can be revoked individually.

\boldtitle{Fallback strategy.}
Since the security of a fallback strategy is equal to the weakest allowed option, L2TP/IPsec with the known pre-shared key and PPTP with its cryptographic weaknesses should be disabled in any automatic protocol selection process. 

As mentioned earlier, one of the main reasons why commercial VPN services provide fallback options is to bypass firewall filters. The fallback strategy for firewall traversal gives an advantage to any malicious access-network operator or oppressive government that wants to attack VPN users. An alternative approach would be to only try safe firewall traversal techniques such as using server ports that are usually not blocked (e.g.~443). Clearly, the commercial VPN service market has led some providers to maximize service availability at the cost of security even when the two are in direct conflict. 

When using RRAS to create VPN connections on Windows clients, the \texttt{VpnStrategy} setting should never be set to 0, 2, 4, 6 or 8. The reason is that all these values instruct Windows to attempt IKEv2, SSTP, PPTP, and L2TP/IPsec, just in different orders, until one succeeds. In an ideal world, Windows would stop supporting protocols and configuration options with known serious vulnerabilities.  
%!TEX root = ../main.tex

\section{Responsible disclosure}
\label{sec:disclosure}

We have reported all the vulnerabilities that we discussed in this paper to the corresponding VPN service providers. We described the attacks and their impact to them as well as provided suggestions on how to mitigate the attacks (see Section~\ref{sec:solutions}). At the moment of writing, 10 of the tested providers have responded to us. They acknowledged all the problems and have fixed all of them, except the pre-shared key issue in L2TP/IPsec and Cisco IPsec, for which they are still discussing about the solution that we have proposed.

In addition, we reported the optional encryption problem of PPTP to the Microsoft security response team with the hope that they would change the default behavior of Windows VPN clients. While they acknowledged the problem, they are already looking into deprecating PPTP, and thus no immediate fixes will be released.

Similarly, we have contacted the SoftEther team about the default value of the \texttt{CheckServerCert} parameter. 
Even though they acknowledge the problem, they are hesitant to change the default value of the software. According to them, it is equally the responsibility of the commercial VPN providers to learn to use the software properly and to educate their customers with clear guidance.

Regarding \texttt{sstp-client}, we have reported its certificate verification problem to the author but have not received a response, and the project appears abandoned.

%!TEX root = ../main.tex

\section{Discussion}
\label{sec:discussion}

It appears that commercial VPN services compete by providing the maximum number of features, such as different VPN protocols, and maximum ease of use, and that security is a secondary concern for them. For example, the issue of using publicly known pre-shared keys for L2TP/IPsec has been known for years, but it has not been addressed by most commercial VPN services. While this issue could be solved by provisioning certificates to the clients as discussed in Section~\ref{sec:solutions}, that would be an administrative hurdle for the VPN service providers and might scare away non-expert customers. This could be the reason why they have opted to continue the insecure services rather than endanger their business growth.

There are also many entirely unnecessary flaws in the VPN client settings, such as not checking the server certificate or server name. These appear to indicate lack of technical knowledge or security awareness by the service providers. We hope that the current paper will, at least to some extent, increase awareness about the importance of correct VPN configuration among the commercial VPN developers and operators.

We further observed the importance of developer documentation and configuration examples for third-party VPN software components that are used as building blocks in the commercial VPN services. Let us take OpenVPN for a positive example. It has detailed documentation of all the software configuration options as well as best-practice guidelines for building secure systems. Also, the OpenVPN software warns the user about insecure settings. These are probably the reasons why none of the commercial VPN services in our study were found to have insecure OpenVPN configurations.

A more sinister explanation for the configuration weaknesses might be that some access-network firewalls are intentionally configured to permit insecure VPN protocols and applications while blocking ones that are not vulnerable. This encourages commercial VPNs to support vulnerable but apparently more reliable settings and protocols, which can then be spied upon. There is anecdotal evidence, both personal experience and online discussions about traveling in China and other countries, of this kind of practice in regions with strict government surveillance of citizens. For example, PPTP often works while OpenVPN does not, even though PPTP is not technically any more difficult to block. We can only speculate about the exact reason for such selective blocking.

%!TEX root = ../main.tex

\section{Related work}
\label{sec:related}

Related work about security and privacy issues of VPNs can be categorized as follows:

\boldtitle{VPN client exploits. } Fazal et al.~demonstrated that an attacker could penetrate into the VPN tunnel by exploiting clients with a dual-NIC that supports both Wi-Fi and Ethernet~\cite{fazal2004tackling}. Similarly, privilege escalation attacks, e.g. in NordVPN~\cite{CVE-2018-3952} and ProtonVPN~\cite{CVE-2018-4010}, allowed an attacker to gain access over the VPN traffic. Security issues have also been found in proprietary VPN clients and devices from Cisco~\cite{CVE-2014-3393,CVE-2016-6415, CVE-2018-0101}, including remote code execution vulnerabilities.

\boldtitle{Information leaks. }There are have been various user de-anon\-y\-miza\-tion attacks that leak information about the VPN user, as presented in the overview by Appelbaum et al.~\cite{appelbaum2012vpwns}. In a majority of the cases, the VPN reveals user information due to IPv6 traffic and DNS leakage~\cite{perta2015glance}. Such leaks can also occur due to advanced VPN functionality such as port forwarding~\cite{portforwardingleak}, WebRTC~\cite{al2017one}, and web cookie synchronization~\cite{papadopoulos2018exclusive}, which are not strictly related to the VPN tunnel but can defeat some of its goals. Most of these issues have also occurred on mobile-device-based VPNs~\cite{ikram2016analysis, zhang2017oh}. There are online detection tools~\cite{ipleaktest,webrtcleaktest, dnsleaktest} with which the users can check whether their VPN connection is leaking any such information.

\boldtitle{Cryptographic vulnerabilities and design flaws. }
VPN protocols have undergone critical cryptanalysis in the past~\cite{schneier1999cryptanalysis, ferguson2000cryptographic, bhargavan2016practical,adrian2015imperfect}. Among others, MS-CHAPv2 in PPTP is known to have cryptographic weaknesses, and it is possible to break the protocol with exhaustive key search~\cite{marlinspike2012divide,schneier1999cryptanalysis}. Moreover, it has also been shown that breaking PPTP requires less effort when it is used in conjunction with a RADIUS server~\cite{horst2016breaking}. If used against a large number of users and connections, these attacks demand relatively heavy computing resources, whereas the configuration flaw explained in this paper immediately exposes the user traffic.

Another known issue of PPTP is that the configuration packets in the PPP negotiation phase are not authenticated~\cite{mudge1998cryptanalysis}. This means that a network attacker can spoof the packet containing the DNS server's address and effectively force all name resolution to happen through a compromised name server. The misconfiguration of PPTP in this paper brings slightly more benefit to the attacker because it allows the attacker to obtain all network traffic of the victim, not just the traffic that involves DNS. Also, by forcing no encryption, the victim's traffic is visible to everyone, not just the attacker.

Oracle-based attacks on VPN systems can also undermine the security of the tunneled traffic. Nafeez disclosed the \textit{compression oracle attack} on OpenVPN compression algorithms, where an attacker can send cross-domain requests when an HTTP website is tunneled through OpenVPN connections~\cite{voracle2018blackhat}. Felsch et al.~demonstrated that reusing the same key pair across IKEv1 and IKEv2 allows an attacker to bypass authentication as well as perform impersonation attacks~\cite{felsch2018dangers}. They exploited Bleichenbacher oracles in the IKEv1 implementations of four large network equipment manufacturers to break a majority of the handshake variants in IKEv1 and IKEv2. 

Weak pseudo-random number generators (PRNG) in the VPN implementation can also subvert the security of the cryptographic protection. The weaknesses could be due to faulty implementation, such as a hard-coded seed key along with a legacy PRNG~\cite{cohney2018practical}, or an intentional backdoor~\cite{bernstein2016dual}, or an intentionally insecure generator~\cite{checkoway2014practical}. In any case, an attacker may be able to recover the secret keys for the VPN tunnel and intercept the traffic passing through the VPN.

There have been state-sponsored backdoors planted by surveillance programs, for example project BULLRUN~\cite{perlroth2013nsa} and TURMOIL~\cite{turmoilprocessing}, which were specifically designed to target VPN connections~\cite{vpnbackdoorprograms}. While open-source implementations of standard protocols may have backdoors hidden in the code or specification, they at least can be audited by anyone. Proprietary protocols and closed-source VPN implementations, on the other hand, are harder to analyze.

%!TEX root = ../main.tex

\section{Conclusion}
\label{sec:conclusion}
In this work, we analyzed the security of how popular commercial VPN providers set up, or instruct their users to set up, desktop VPN clients. We studied commonly used VPN protocols and software on Windows, macOS, and Ubuntu. We found vulnerabilities in the client configurations of most of the protocols and clients. These vulnerabilities allow network attackers to perform MitM or server impersonation on the connection and thus obtain the victim's original network traffic. Similarly, local attackers can exploit vulnerabilities to steal user credentials for the VPN services. We provide guidelines for fixing these vulnerabilities.

The main message of this paper is that security flaws, either accidental or intentional, are not always deeply hidden in the code or cryptography. Instead, simple configuration mistakes, poor instructions, insecure default values, and failure to disable broken legacy features can result in widespread security failures across an entire industry.

\bibliographystyle{ACM-Reference-Format}
\bibliography{references}
\end{document}